# Spatial polarisation within foreign trade and transnational firms' networks. The Case of Central and Eastern Europe

Natalia Zdanowska

Centre for Advanced Spatial Analysis, University College London,
UMR 8504 Géographie-cités, Université Paris 1 Panthéon-Sorbonne
n.zdanowska@ucl.ac.uk

**Abstract**

After the fall of the Berlin Wall, Central and Eastern Europe were subject to strong polarisation processes. This article proposes examines two neglected aspects regarding the transition period: a comparative static assessment of foreign trade since 1967 until 2012 and a city-centred analysis of transnational companies in 2013. Results show a growing economic differentiation between the North-West and South-East as well as a division between large metropolises and other cities. These findings may complement the targeting of specific regional strategies such as those conceived within the Cohesion policy of the European Union.

**Introduction**

Globalisation is characterized by strong spatial and economic polarisation processes as well as competition to capture the most innovative technologies (Held McGrew Goldblatt et al. 1999; Greenstock 2007). Polarisation is 'the attraction exerted by a place on a more or less extended and heterogeneous one that is in a situation of dependency with respect to this centre' (Elissalde 2004, 1). The attraction of these 'growth poles' is a driving force of the development of a regional whole, where selective investments are creating growth-multiplying mechanisms (Perroux 1955). Generally examined at the level of countries or regions (Krugman 1995), these effects can as well captured when considering cities as they are relevant actors of globalisation (Alderson and Beckfield 2004; Beaverstock et al. 1999; Derudder 2006; Sassen 1991; Taylor 2003). At the cross-over of the major international networks and flows (Rozenblat and Pumain 1993; Batty 2013), they are the major receptors of economic activity attracting transnational companies and foreign direct investments (Massey 2007; Berger 2005; Markusen 1994; Finance and Cottineau 2018; Zdanowska 2018).

The effect of globalisation processes can be particularly observed in the evolution of Central and Eastern European countries (CEEC) and cities (CEECc)[1] economic specialisations. Central and Eastern Europe (CEE) integrated into the market economy in the early 1990s, when strong hierarchical forces were already active between cities in the rest of Europe and the world within the globalized markets (Rozenblat and Pumain 2007). The contribution of this article is to evaluate the effect of intensification of international trade and foreign ownership of firms' capital on polarisation in CEE as a result of entering of CEEC and CEECc into the globalized processes of market economy.

Studies have shown that at country level foreign trade is regarded as a source of economic growth and regional development (Makki and Somwaru 2004). The attraction of transnational companies in cities (Rozenblat and Pumain 2007) is an opportunity for redistribution of wealth in a country (Makki and Somwaru 2004; Amin and Thrift 1994; Andreff 1996). Additionally, economic specialisation in high-intensive technology and knowledge-based services is considered as the highest stage of integration in the globalisation processes and the global production networks (Krugman and Obstfeld 2000; Massey 2007; Finance and Cottineau 2018).

---

[1] Understood as eight post-communist countries, members of the European Union (Bulgaria, Croatia, Czechia, Hungary, Poland, Romania, Slovakia, Slovenia). CEE will be used in the rest of the article as an abbreviation for Central and Eastern Europe, CEECs for Central and Eastern European countries

The economic transition post-1989 and accession of the CEECs to the European Union since 2004 have been major stimuli to intensification of foreign trade and capital flows into the CEE economies in the forms of foreign direct investment (FDI),[2] portfolio investments and loans (Pyka 2011).

In spite of the general idea that inflow of FDI is an essential determinant of economic growth and a wealth and job generator (Barrell and Pain 1997; Olofsdotter 1998; Berger 2005; Chowdhury and Mavrotas 2006; Shi et al. 2018) the impact on the host countries can be variable. Some works have shown that it may contribute to deepen inequalities already present between countries and cities (Ram and Honglin Zhang 2002; Lipsey and Sjöholm 2005; Zdanowska 2017). In CEE, immediately following the fall of the Berlin Wall, CEECs were characterised by a centralisation of all economic and administrative management centres in the capital city, as a reflection of the planned economy approach (Śleszyński 2002). Notwithstanding anticipated tendencies for spatial structure to deconcentrate following reforms of the administrative division, firms' headquarters remained largely located in capitals and other metropolitan areas throughout the 1990s and 2000s (Guzik and Gwosdz 2000). There remains no doubt as to the major role metropolises play in attracting foreign direct investment, as they concentrate major economic functions (Sassen 1991; Taylor 2004; Vandermotten et al. 2010) and attract the most qualified and dynamic activities (Rozenblat and Pumain 2007).

For this purpose, the economic differentiation within CEE will be examined in the first part of this paper concentrating on the relatively neglected consideration of the years before and after the fall of the Berlin Wall. Studies have been conducted on regional sectoral change in CEE between 1995 and 2004 (Capello Perucca 2017), but scarce harmonised and longitudinal data, make the years before the fall of the Berlin Wall difficult analyse. In this work, the evolution of each CEEC trade specialisation, as well as the trade flows between these countries are analysed from 1967 to 2012. The following questions are posed. Have CEECs evolved towards similarity in exported products? Did the advent of the market economy stimulate economic differentiation?

---

[2] Foreign direct investment is a "cross-border investment by a resident entity in one economy with the objective of obtaining a lasting interest in an enterprise resident in another economy" (OECD 2005, p. 50). The main difference between FDI and portfolio investment is that, in the latter case, an investor purchases equities in foreign companies, while in the former an investment leads to substantial influence or effective control of the decision-making process in a foreign company. The basic criterion representing an investor's influence on the management of an enterprise is ownership of at least 10% of the voting power of the enterprise, while 50% or greater ownership implies complete power in terms of decisions (OECD, 2005). The latter criterion will be considered in the article as a whole.

In the second part of the article the focus is on multinational company networks in 2013 at the city-level. The latter is particularly relevant as in CEE the regional GDP is very unbalanced in each country. Moreover, this analysis has not previously been undertaken for all the cities, regardless of size, in the eight CEE countries. Other studies on FDI and transnational companies have focussed mainly on the capital cities, or on the regional and national levels only. This last section aims to evaluate the metropolitan effect in CEE in 2013 considering the economic specialisation of all cities as determined by the activity sectors of firms controlled by transnational companies. The role of large, medium-sized and small cities will be examined.

This work should contribute to broader knowledge on the spatial polarisation effect in globalized networks and inform and complement the targeting of specific regional strategies conceived in the Cohesion policy of the European Union under the heading Smart Specialisation (S3) which seeks to prioritise domains, areas and economic activities where regions have a competitive advantage and the potential to generate knowledge-driven growth (European Union, 2011).

**Materials and Methods**

To study the economic differentiation between countries within international trade networks and to test the hypothesis of spatial polarisation and metropolises effect in Central Eastern European urban systems within transnational firms' networks, we use three different sources:

- Trade flows for each year between 1967 and 2012, detailed by exporting and importing country and traded product categories from the *CHELEM* database (see Zdanowska 2018);[3]

- Population of cities in 2011 defined as urban agglomerations for all European cities in the *TRADEVE* database (for more see Bretagnolle et al. 2016). A distinction has been made between large, medium-sized and small cities in CEE [see Appendix 1];

- Capital control links of firms in 2013 at city level from the *ORBIS* database.[4] It lists all the companies, located outside CEE, possessing capital of CEE companies in all types of sectors. Additionally, information about CEE companies controlling the capital of other firms in CEE, but also in other European Union countries, is also included (see Zdanowska 2018). These variables permitted the reconstruction of capital control links between cities where companies are localized and to understand which CEE cities are the most concerned by these links.

The main methods used for analysis in this article are statistical correlations, factor analysis and a gravity model comparison to understand the evolution of spatial interactions and the characteristics of CEE countries and cities.

---

[3] The *CHELEM* database (Harmonised Accounts on Trade and the World Economy), developed by the Centre for Prospective Studies and International Information (CEPII) in Paris, provides data on international trade. *CHELEM* is built and actualized from the United Nations *COMTRADE* database (De Saint Vaultry 2008). *CHELEM* provides access to harmonized annual trade data harmonized for over 45 years. Uniquely, it allows one to trace the evolution of main exchanges of each Central and Eastern European countries with other countries, before and after the fall of the Berlin Wall. At the same time, it allows one to observe the evolution of the sectors of products exchanged over this long period for each country and determine the changes in specialisations over time. *CHELEM* gives a relatively comprehensive view of world trade, with homogeneous time and space series (De Saint Vaultry 2008). When some countries fail to report their trade or do it with delay, and product codes evolve over time and at different times depending on the country, CEPII corrects the missing flows by estimation, and harmonizes the declarations by exporters and importers to avoid repeated information.

[4] The *ORBIS* database produced by the Bureau Van Dijk (BVD) is a global database on capital links between mother and daughter companies. The extraction of *ORBIS*, made available for this article, has been made upon a sample of the original database were a selection of the 3 000 biggest parent companies has already been applied, which reduces the number of possible subsidiaries. The uniqueness of *ORBIS* is to make available at city level, data on both companies controlling capital and on the companies they own, for the eight CEEc considered.

*Correspondence factor analysis and hierarchical bottom-up classification*

The evolution of the main products exchanged by the CEEC was examined in terms of their export trajectories regarding sectors, between 1970 and 2010 (Pumain and Saint-Julien 2001). For this purpose, a correspondence factor analysis, followed by a hierarchical bottom-up classification was carried out using *ca* and *ade4* packages in R software (Nenadic and Greenacre 2007; Dray and Dufour 2007). The data covers nine different years,[5] ten main sector groups[6] and the corresponding exported volumes. The export sectors were aggregated into 10 major groups. The first and second factorial axis summarize respectively 55.72% and 13.67% of the information, which is sufficient for the analysis as it represents together nearly 70% of the information. The specialisation trajectory of each country was then projected onto the factorial plan.

Similarly, to analyse the economic specialisation of CEE cities concerned by foreign capital control links, a correspondence factor analysis has been conducted based on a matrix of CEE cities, in rows, and the respective economic specialisations of the companies in columns. An aggregation into the 9 most representative foreign capital control sectors of firms in CEE has been constructed.[7]

*Gravity model and Poisson regression*

We modelled the trade fluxes by a simple formalisation of a gravity model. The gravity model assumes that, 'exchanges between two regions or two cities will be more important as the weight of cities or regions is large and that the distance between them is low' (Pumain 2004, 1). We have considered the respective GDP of each pair of East-Central European countries,

---

[5] 1970, 1975, 1980, 1985, 1990, 1995, 2000, 2005, 2010.
[6] Agriculture, Chemistry, Construction, Energy, Food Products, Mechanics, Mining, Siderurgy, Textiles, Wood.
[7] The sectors are the following*:* car industry *(*repair and sale of motor vehicles); finance/insurance/banking (life insurance, financial leasing); IT (IT services activities); industry (industrial production of chemical products, cement, textiles, plastics, household appliances, paper); media/advertising/communication (organization of the television program, radio, wireless telecommunications activities); real estate/tourism (real estate agency, hotels and similar housing); sales/trade (sale of machinery, chemicals, pharmaceuticals, textiles and food products); services/construction; energy.

the exchanges between them and the distance separating their capitals [see Appendix 2].

This model was estimated via a generalized linear Poisson regression, the performance of which is much better than a simple regression (Flowerdew & Aitkin, 1982). The relevance of this statistical approach has also been verified for gravity models in geography (De Aubigny et al. 2000).[8]

In practice, the estimation was realized by the *glm2* (generalized linear model) package in R software (Marschner 2011).

**Table 1. Estimation parameters of GDP *(β et γ) and distance (δ)* variables of the regression and coefficients of determination in 1967, 1992, 2002 and 2012[9]**

|     | 1967 | 1992 | 2002 | 2012 |
|-----|------|------|------|------|
| β   | 0,8  | 0,5  | 0,8  | 1,2  |
| γ   | 0,8  | 0,5  | 0,8  | 1,1  |
| δ   | 0,4  | -1,7 | -1,1 | -1,8 |
| R²  | 0,6  | 0,6  | 0,7  | 0,8  |

The variable to be explained is the observed trade flux and the explanatory variables are the GDP (*Mi* and *Mj*) and the distance (*Dij*). The following formula describes the regression with the error term $u_{ij}$, following a Fish distribution:[10]

$$\log Fij = \alpha \log k + \beta \log Mi + \gamma \log Mj + a\, \delta \log Dij + u_{ij}$$

**Results**

*Widening West-East division*

All CEEC are characterized by a clear change in the nature of their main exports, starting from 1990, as a direct impact of the shift to the market economy. Three types of trends can be identified.

---

[8] We also do not encounter the problem of a large number of zero flows for which the estimator should be adapted (Martin and Pham 2015).
[9] The confidence interval is 95%.
[10] $k$ is a constant not relevant for the following interpretations and $a$ is a coefficient fixed for the purposes of the analysis and equal to 2.

Bulgaria and Croatia – specializing in agriculture, the production of energy (extraction of coal, oil and gas) and ores (iron) – have turned, after 1990, towards exports of textile products (clothing, leather). However, in 2010, they returned to their communist period specialisation.

**Figure 1. Specialisation trajectories of each CEEC's world export sectors between 1970 and 2010**

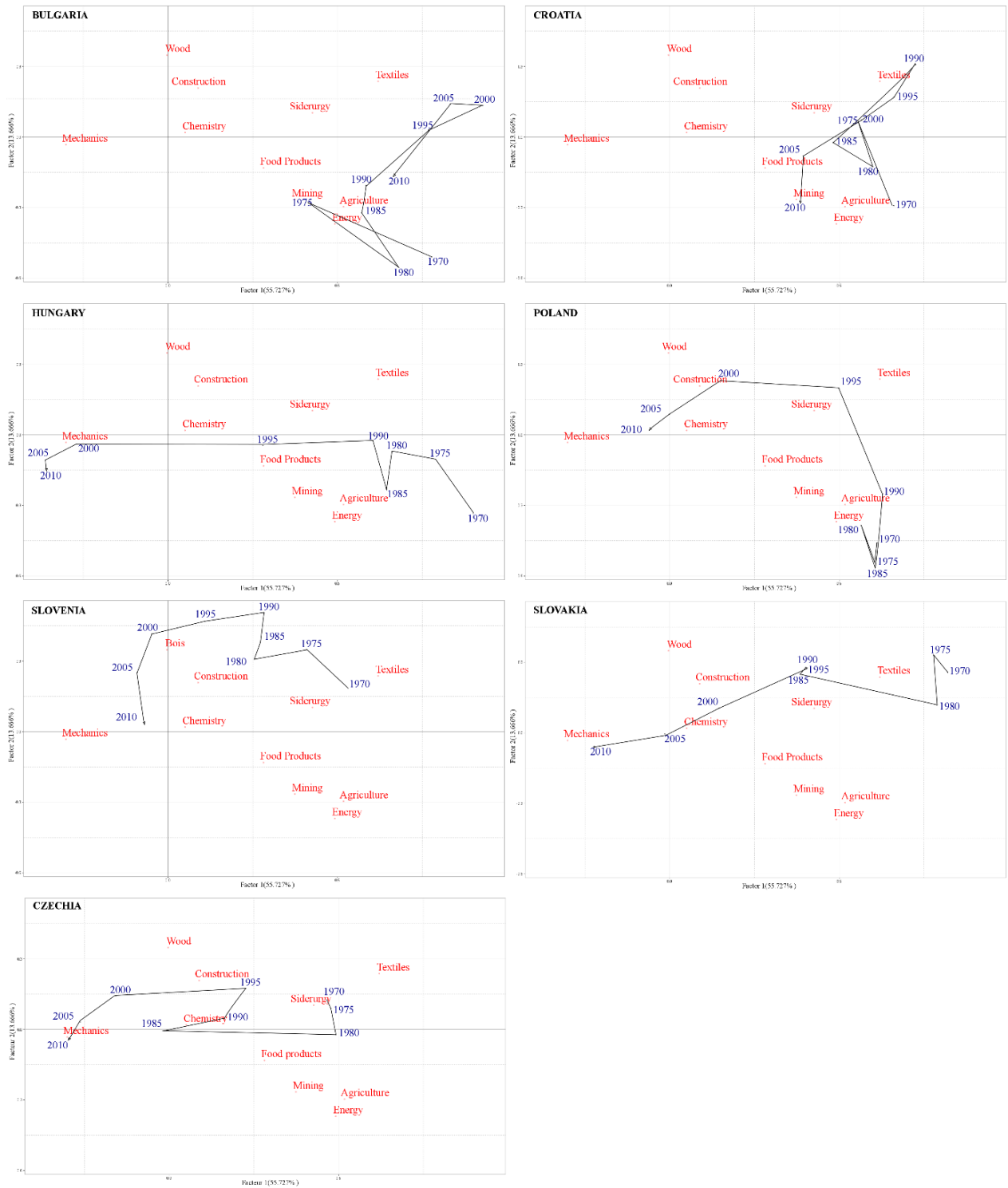

In fact, the textile industry was of great importance in Bulgaria, due to its historical background harking back to the Ottoman Empire, which could explain this continuity (Laferrère

1989).

Bulgaria and Croatia have thus failed to diversify further, despite their attempts since 1990. This can be explained in particular by the important oil port activity in Burgas and Rijeka (Kutsarov et al. 2007).

On the other hand, Poland and Hungary – which were also exporters in the field of agriculture and energy during the Communist period – have at the same time experienced an exceptional rise in their range of exported products. Poland specialized in steel and construction in the years 2000, then in 2010 in chemistry (pharmaceuticals) and electronics and mechanics (motor vehicles, computer equipment). The latter specialisation, under high technology, was reached as early as 2000 by Hungary, after a period dominated by the exports of food products in the 1990s.

Moreover, Slovenia, Slovakia and the Czechia – already specialized in textile products during the communist era (steel in the Czech case) – converted extremely rapidly to construction products from 1990, and then to mechanics (pharmaceuticals in Slovenia) after 2005.

These analyses show the significant effects of the reorientation of trade in the 1990s on the diversification of the commercial structure of the CEEC. Croatia and Bulgaria aside, all countries have experienced a rise in the range of exported products, by achieving specialisation in high technology in some cases. Hungary, for example, specialized by 2010 in the automotive industry despite its long-standing agricultural tradition (Kiss 2004). This diversification also reflects strong economic polarisation processes between CEEC as Croatia and Bulgaria have not evolved in the same way and in the same range of products as Poland or Slovakia for example (Lončar and Braičić 2016).

These polarisation processes can be confirmed in terms of interactions between CEEC over time. The most important trade flows have taken place between the Western countries of the region (Poland, Czechia, Hungary and Slovakia) over the entire period.

**Figure 2. Export flows within CEE between 1967, 1992 and 2012, expressed in millions of 2012 constant dollars**

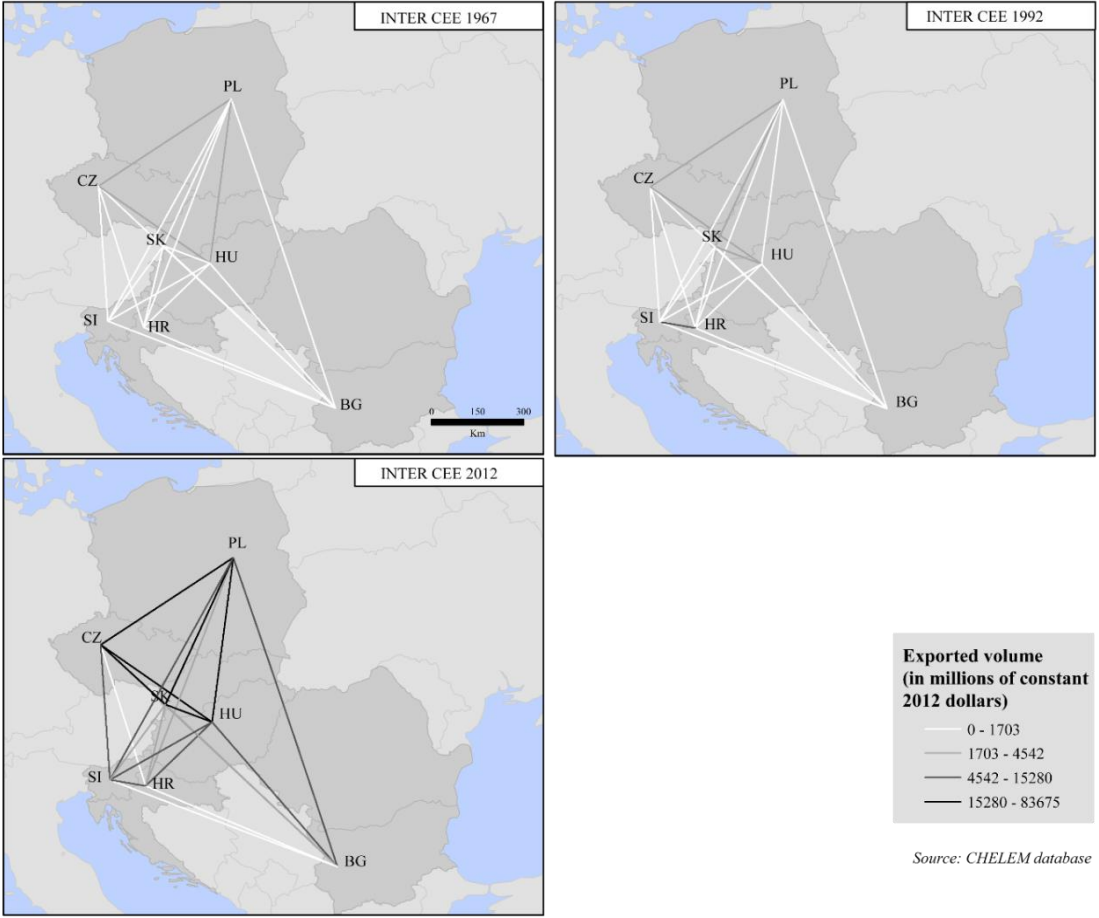

The main increase in volume exchanged occurred in 2012. The majority of exports came from Czechia and then from Poland. In contrast, the share of exports from Bulgaria, Croatia and Slovenia was very low. It seems that there is a more integrated region in CEE concentrated in the North-West. Two major factors that seem to determine the volume exchanged are the economic weight of the countries and the distance between them.

In fact, the results of the regression show that the distance coefficient δ decreased between 1967 and 2012, which means that the distance was an increasing constraint for the establishment of trade flows between CEEC. Moreover, the influence of GDP has increased in 2012 as shown by the increase of coefficients $β$ and γ. Finally, the quality of the relationship has expanded over time – especially after the 2000s (from $R^2$ = 0.6 in 1967 to $R^2$ = 0.8 in 2012). The flows thus became more polarized on the most developed and closest countries, which explains the important trade between Poland, Czechia, Hungary and Slovakia in 2012 observed earlier. This confirms a regionalization effect in CEE around these four countries over time, which was enhanced by the entrance to the market economy and the globalization.

When considering capital control links between CEE companies in 2013 expressed firstly at national level, the hypothesis of flows polarisation and West-East disequilibrium is confirmed. Hungary is responsible for the largest amount of firms' investments in other CEE countries (44% of total revenues generated), followed by Poland (21%) and Czechia (14%).

**Table 2. Amount of FDI revenues generated by companies from other CEE countries, in millions of Euros in 2013**

|        |    | Destination |     |     |       |      |     |       |       |     |
|--------|----|-------------|-----|-----|-------|------|-----|-------|-------|-----|
|        |    | CZ          | PL  | HU  | SK    | HR   | SI  | BG    | RO    | %   |
| Origin | CZ |             | 52  | 1   | 383   | 0,4  | 0,1 | 215   | 203   | 14  |
|        | PL | 1 148       |     | 9   | 28    | 9    | 1   | 0,4   | 73    | 21  |
|        | HU | 30          | 31  |     | 1 676 | 607  | 60  | 96    | 235   | 44  |
|        | SK | 390         | 194 | 33  |       | 0,6  | 10  | 1     | 0,1   | 10  |
|        | HR | 8           | 8   | 2   | 28    |      | 128 | 0,01  | 0,1   | 3   |
|        | SI | 8           | 2   | -   | 0,3   | 212  |     | 1     | 23    | 4   |
|        | BG | 1           | -   | -   | -     | 0,05 | -   |       | 3     | 0   |
|        | RO | -           | -   | -   | -     | -    | -   | 953,8 |       | 4   |
|        | %  | 26          | 5   | 1   | 34    | 13   | 3   | 9     | 9     | 100 |

Bulgarian, Romanian and Croatian companies are the ones investing the least in other CEE countries. The only capital link from Romania is from Bucharest and directed towards Sofia in Bulgaria. Hungary is at the origin of the most important FDI incomes and is represented in the most countries, probably due to its central geographical location in the Central-Eastern European area.

At city level, these interactions between companies within the Central Eastern European area, also appear very asymmetrical.

**Figure 3. Capital control links between Central-East European companies in 2013 aggregated at the city level**

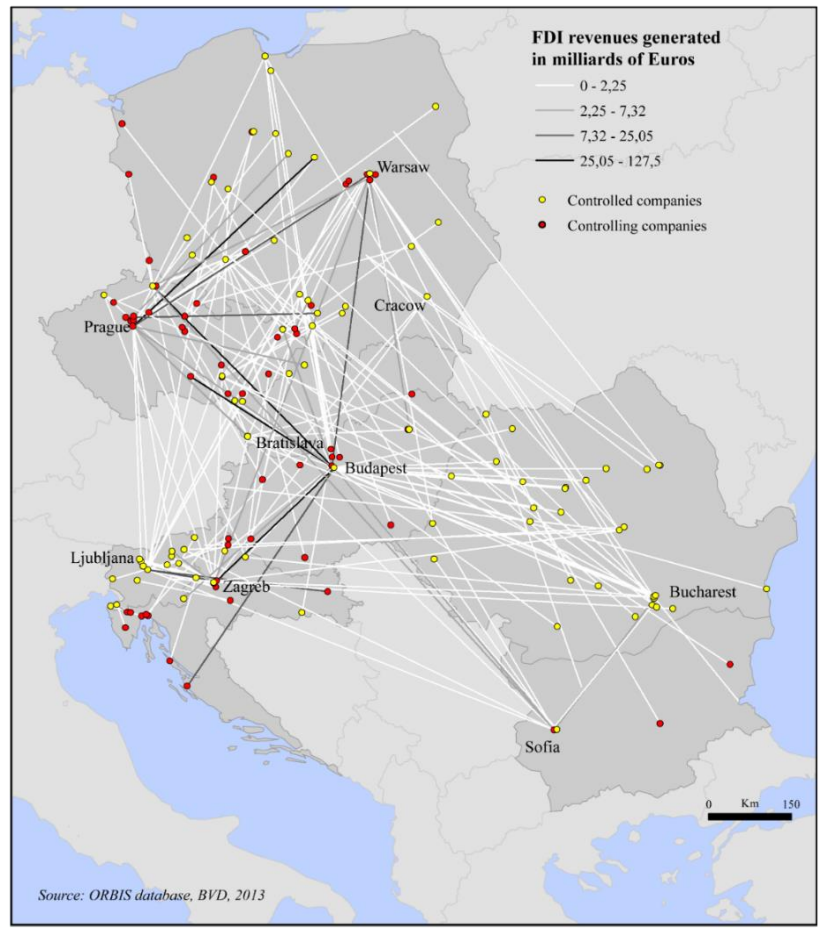

The strongest control links connect Prague, Warsaw and Budapest and reveal a clear regionalization effect in West-Central Europe leaving the Eastern cities of the region out of the strongest interactions scheme. These results confirm our observations at country level of trade flows and our hypothesis of strong spatial polarisation effects as a result of entering into globalisation processes.

*Metropolises vs other cities opposition*

In this section, we tried to understand if polarisations identified at CEEC and CEEc interaction level drive to a differentiation of the economic specialisation of CEE cities themselves.

**Figure 4. CEE cities and their economic specialisations in multi-level links in N-1**

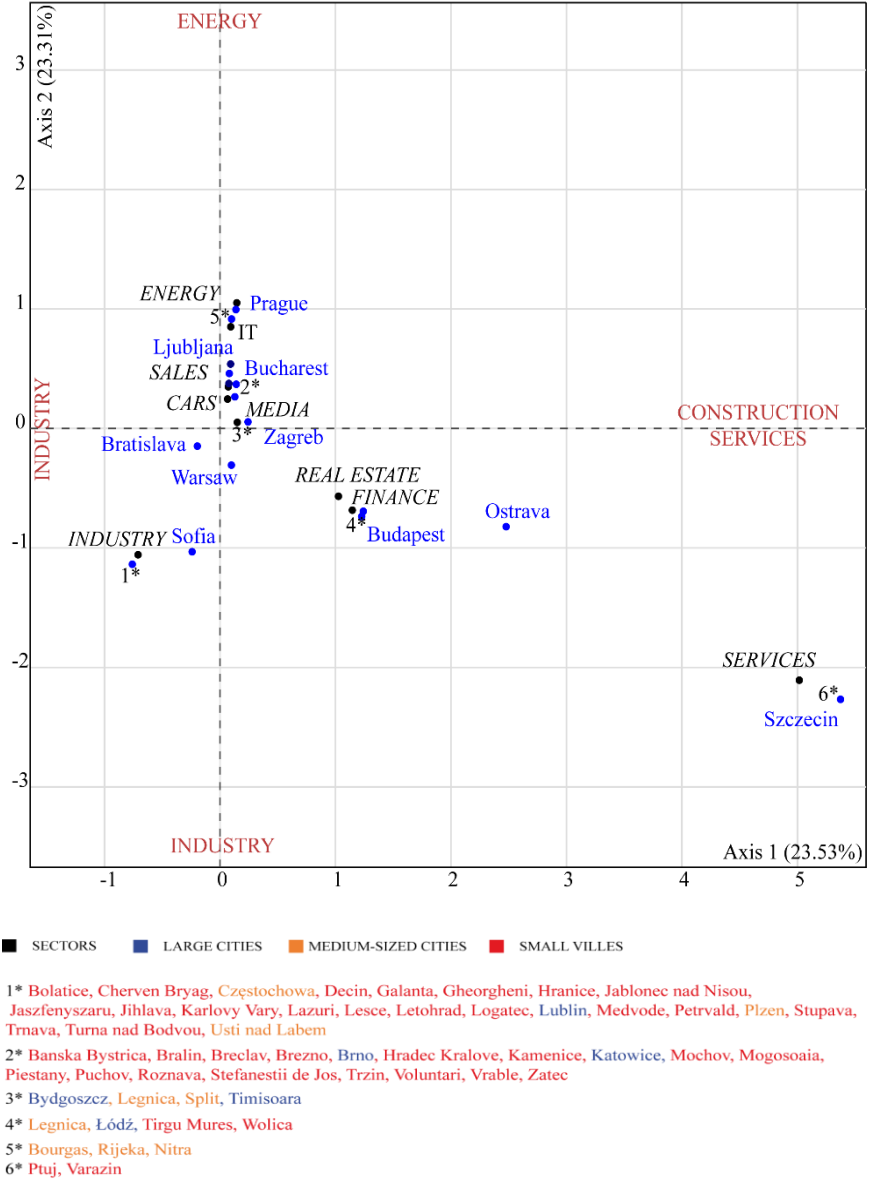

1* Bolatice, Cherven Bryag, Częstochowa, Decin, Galanta, Gheorgheni, Hranice, Jablonec nad Nisou, Jaszfenyszaru, Jihlava, Karlovy Vary, Lazuri, Lesce, Letohrad, Logatec, Lublin, Medvode, Petrvald, Plzen, Stupava, Trnava, Turna nad Bodvou, Usti nad Labem
2* Banska Bystrica, Bralin, Breclav, Brezno, Brno, Hradec Kralove, Kamenice, Katowice, Mochov, Mogosoaia, Piestany, Puchov, Roznava, Stefanestii de Jos, Trzin, Voluntari, Vrable, Zatec
3* Bydgoszcz, Legnica, Split, Timisoara
4* Legnica, Łódź, Tirgu Mures, Wolica
5* Bourgas, Rijeka, Nitra
6* Ptuj, Varazin

The most important observation is the existence of certain cities, which are the headquarters of transnational companies of high-technology intensive sectors, such as Prague, while others, such as Sofia, are rather concentrated on low technology. Other cities, mainly small as Bolatice, Cherven Bryag, Decin or Galanta (except Sofia), are specialized only in single industries, while Warsaw or Budapest are characterized by companies from several sectors.

The more detailed analysis of the three first factorial analysis revealed strong oppositions between cities confirming the specificity of the ones from the former industrial Silesian basin.

The coordinates corresponding to the first factorial axis oppose cities specializing in construction services (positive coordinates in red) and specialized cities in the industry (negative coordinates in blue).[11]

**Figure 5. Values of city coordinates in CEE of the first factorial axis**

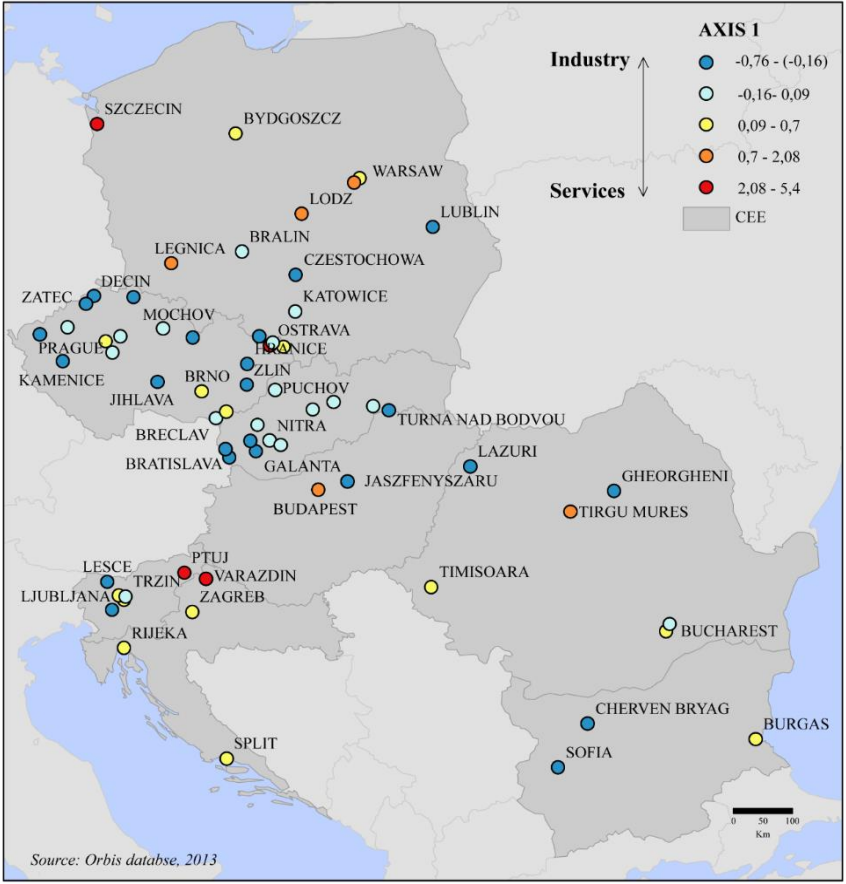

The industrial towns of the Silesian Basin are opposed to Ptuj, Varazin and Szczecin specialized in transport services and construction. This result still shows today, evidencing a certain continuity of the industrial character of the Silesian Basin, converted since 2010 to the production of engines and motor vehicles. Cities such as Zlín, Katowice, Legnica or Decin are the targets of the NOVUM and TRITRIA programmes within their respective European

---

[11] It is necessary to analyse the first three factorial axes, as according to the histogram of proper values they together represent 65.3% of the total inertia (the axis 1 represents 23.53%, the axis 2, 23.31% and axis 3, 18.42%).

Groupings of Territorial Cooperation. They were established to facilitate and support Polish-Czech border cooperation and strengthen economic and social cohesion of their area. One of the objectives of these projects is to support an energy industry research centre.

The cities of Ptuj and Varazin, located in the more industrial North of Croatia, maintain their industrial services specialisations from the past (Lončar Braičić 2016). Similarly, Szczecin, located not far from the Baltic Sea with an access to it via a channel, has always been characterized by maritime and port construction activities. It is an important node for transporting European goods between Germany and Scandinavia and is a node of the Central European Transport Corridor supporting the policy of the European Grouping Territorial Cooperation.

The second factorial axis opposes cities' specialisations in energy (positive coordinates in red) and in construction services (negative coordinates in blue).

**Figure 6. Values of city coordinates in CEE of the second factorial axis**

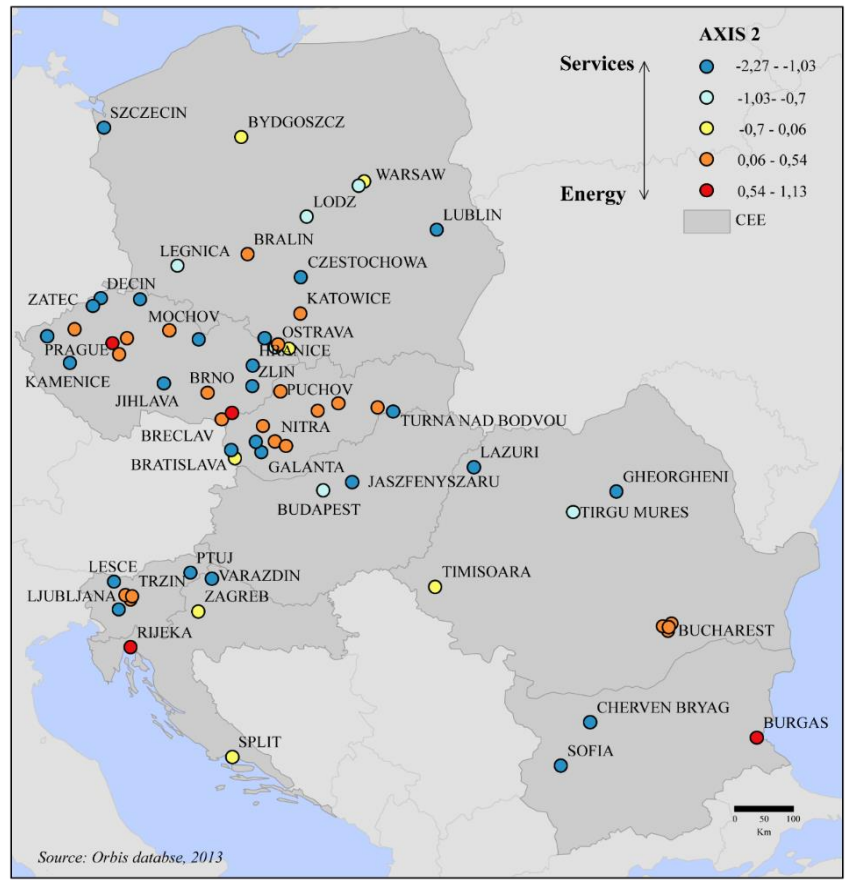

At one extreme, the cities of Burgas and Rijeka can be identified. These are industrial ports and large oil refineries in Bulgaria and Croatia, emblematic of the energy sector. Burgas is located on strategic supply paths necessary for Western Europe, through the 'southern Corridor' of gas pipelines (Glamotchak 2014). The city is the seat of the largest oil refinery in southern Europe, Lukoil Neftochim, controlled by the Russian giant Lukoil (Kutsarov et al. 2007). Rijeka has been a strategic Croatian port since the Yugoslav period, during which the city delivered 10% of Yugoslavian exports. Today, Rijeka's port hosts a large oil refinery, operated by INA and belonging to the Hungarian group MOL. Lukoil is also present and is a testament to the strategy of establishing Russian companies in the Black Sea (Glamotchak 2014). These results coincide with our observations in the first section regarding specialisations of Bulgarian and Croatian exports, mainly oriented in 2010 towards the energy sector, as previous to 1990. At the other extreme, a group of cities can be identified in Southern Poland, Eastern and Northern Czechia (Ostrava, Brno, Zatec), and also the cities of Bratislava, Lazuri and Gheorgheni (Romania), Sofia and Cherven Bryag (Bulgaria). They are all very specialized in the services sector, in particular in companies related to the construction of residential and non-residential buildings, as well as roads.

Finally, the analysis of the coordinates of the cities of the third factorial axis reveals the existence of an opposition between cities specialising in construction services (positive coordinates in red) and cities specialising in finance (negative coordinates in blue).

**Figure 7. Values of city coordinates in CEE of the third factorial axis**

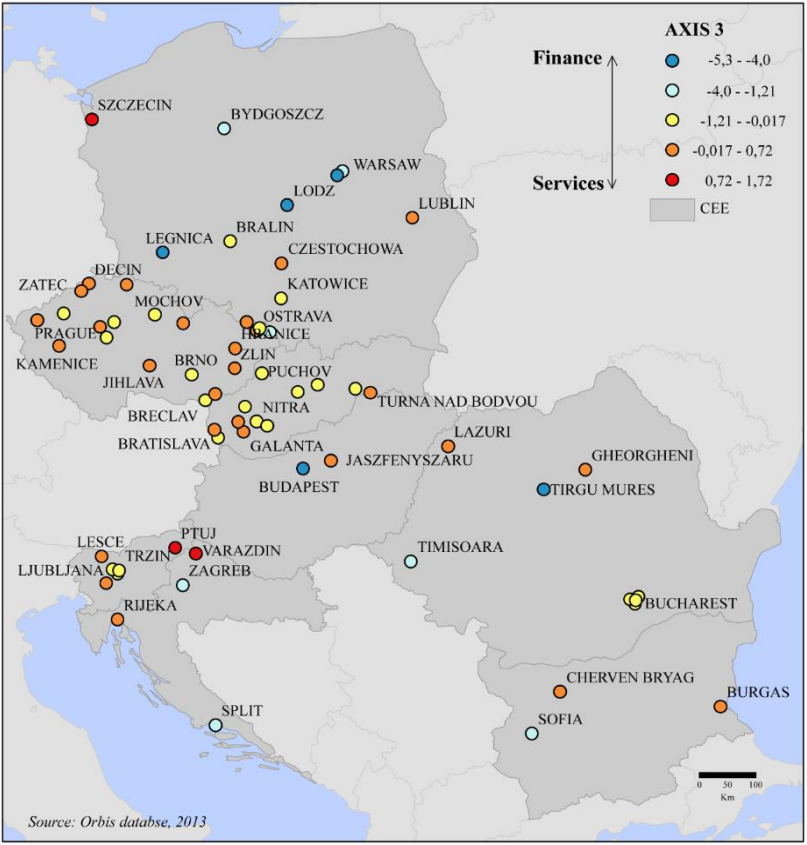

The cities of Ptuj, Varazin and Szczecin – already identified previously as specialized in services, are opposed to Warsaw, Łódź, Budapest, and Tirgu Mures – specializing in finance. This opposition of the specialisations of cities between the finance and services sectors has also been revealed in the case of France (Paulus 2004). CEE's specificity though, is the prevalence of low intensive type services – related to construction.

To sum up, four groups of highly specialized cities, differing in size, could usefully be a target of the European Union regional policies:

• Warsaw, Łódź and Budapest – large/capital cities specializing in finance;

• Burgas and Rijeka, medium-sized cities in the energy sector;

• Ptuj, Varazin (small cities) and Szczecin (large) hosting construction services companies;

• Small and medium-sized cities and towns in Silesia and Northern Czechia (Zatec), characterized by the presence of transnational companies in the industry sector.

In addition, the empirical results have shown that firms in the automotive sector are mainly located in large cities of more than 250 000 inhabitants (72% of the cases). The same goes for the media sector (75%) and real estate (100%) (Table 3).

**Table 3. Proportion of small, medium and large-sized cities with regards to economic sectors at N-2 multilevel control links (in %)**

|             | SMALL | MEDIUM | LARGE | TOTAL |
|-------------|-------|--------|-------|-------|
| CARS        | 14    | 14     | **72** | 100   |
| FINANCE     | 9     | 18     | **73** | 100   |
| IT          | 0     | **50** | **50** | 100   |
| INDUSTRY    | **53** | 20    | 27    | 100   |
| MEDIA       | 8     | 17     | **75** | 100   |
| REAL ESTATE | 0     | 0      | **100** | 100  |
| SALES       | **50** | 21    | 29    | 100   |
| SERVICES    | 29    | 14     | **57** | 100   |
| ENERGY      | 33,3  | 33,3   | 33,3  | 100   |

Companies in the IT sector are located exclusively in large and medium-sized cities, according to an identical distribution. Firms in the industrial sector are, in more than half of the cases, located in small cities with less than 50 000 inhabitants (53%), and in 20% of the cases, in medium-sized cities (between 50 000 and 250 000 inhabitants). The same situation obtains in the sales sector (50% of small cities, 21% of average ones). A clear size effect can therefore be identified, ranging from the most innovative services sectors in the case of the large cities to the less-intensive type of industry in the case of the smallest cities. Finally, this analysis reveals that in 73% of cases large cities are mainly pluri-specialized, as well as some average-sized cities (27%), while small cities are mono-specialized in 65% of the cases in our sample.

**Table 4. Size of cities and sector diversity, given in % of the cases sampled**

|  | SMALL | MEDIUM-SIZED | LARGE | TOTAL |
| --- | --- | --- | --- | --- |
| MONO-SPECIALISATION | 65 | 24 | 11 | 100 |
| PLURI-SPECIALISATION | 0 | 27 | 73 | 100 |

However, it is worth adding that small cities represent more than half of the cities involved in capital links in terms of number. They are the most specialized among cities in CEE, although some large cities are mono-specialized (Sofia, Ostrava).

The mono-specialisation of cities, into one sector, is the result of the development of a cycle of innovations having 'specifically selected a particular group of cities by specializing them in relation to the rest of the urban system' (Bretagnolle and Pumain 2010, 8). In CEE, this was the case during the industrial revolution regarding small towns in the former mining and steel basins of Silesia, but also of the major industrial cities whose development began under communism. Pluri-specialized cities are most frequently capital cities. These pluri-specialized cities were mainly those that gave rise to the new technologies of the nineteenth century, stimulating the creation of high-level services, particularly in the automotive sector (Bohan 2016).

Therefore, all of these results bring evidence of an uncertain effect of FDI attraction in CEE as they revealed strengthening polarisation effects between large metropolises on one side and medium and small-sized cities on the other side as a clear consequence of metropolisation. It is worth remarking, that compared to a country-level approach, a city-level analysis permits a more balanced representation not overemphasising the role of the richest countries in CEE, as medium-sized cities in Bulgaria (Burgas) or Croatia (Rijeka), also play an important role in attraction of foreign firms (Glamotchak 2014).

**Discussion**

This article aimed at investigating the effect of the fall of the Berlin Wall and CEE's subsequent entry into the market economy, on the spatial economic polarisation processes in Central and Eastern Europe. Two original aspects were considered regarding time and space: the evolution of the trade flows and exchanged products of Central and Eastern European countries between 1967 and 2012 and the capital control links between firms in CEE at city level in 2013.

The results have shown strong differentiation and polarisation in the region through time, concentrated in Poland, Hungary, Czechia and Slovakia. Czechia and Slovakia, in particular, have evolved towards high technology trade specialisations in the automotive industry and mechanics. As for Bulgaria and Croatia, they initially specialized in textile production following the communism collapse, returning back to energy in the 2000s as was the case before 1990. Poland, Hungary Czechia and Slovakia are also the countries that exchange the most between each other since 1967. A gravity model analysis has shown that these exchanges have concentrated in the closest and richest countries, over time, confirming a regionalisation effect and a West-East divide in CEE.

The factorial analysis of cities involved in capital links between companies confirmed this polarisation today at city level. It seems that capital and size effect matters in CEE regarding specialisation in high-technology intensive FDI investments and in different types of sectors at the same time. A selection of four types of cities has been revealed: Warsaw, Łódź and Budapest, large capital cities specializing rather in finance; Burgas and Rijeka, medium-sized cities characterized by the energy sector; Ptuj, Varazin (small cities) and Szczecin (large city), hosting construction services companies and small and medium-sized towns in Silesia and Northern Czechia (Zatec), characterized by the presence of transnational companies in the industry sector.

All these differentiations at country and city level drive to a conclusion upon a dual economic and spatial development in CEE since the integration to the globalisation processes between the North Western (Poland, Hungary, Slovakia, Czechia) and the Eastern part (Romania, Bulgaria, Slovenia and Croatia). In addition, a strong differentiation between big metropolises and other cities in CEE has taken place, as a consequence of entering into the neoliberal economic system.

These results open a discussion upon the impact of foreign trade strengthening and FDI attraction on hosting countries and cities. It questions the future development of CEE as a region with a strong heritage of economic relations from the past. It also provides relevant policy considerations regarding the reduction of the economic gap between European Union cities.

In fact, the results on growing differentiation between capital cities, medium-sized and small cities in terms of economic specialisation, may inform and complement the targeting of specific regional Smart Specialisation Strategies (S3) conceived within the Cohesion policy of the European Union. The latter prioritises domains, areas and economic activities where regions have a competitive advantage and the potential to generate knowledge-driven growth. The contribution of this article is to identify these priority areas at city-level and to highlight the economic potential of mid-sized cities in achieving strong city regions – an important policy goal. Indeed, the importance of small and medium-sized cities "should not be underestimated as they play a pivotal role within regional economies" (European Union 2011, 4).

# Appendix

## Appendix 1. Large, medium and small cities in CEE in 2011

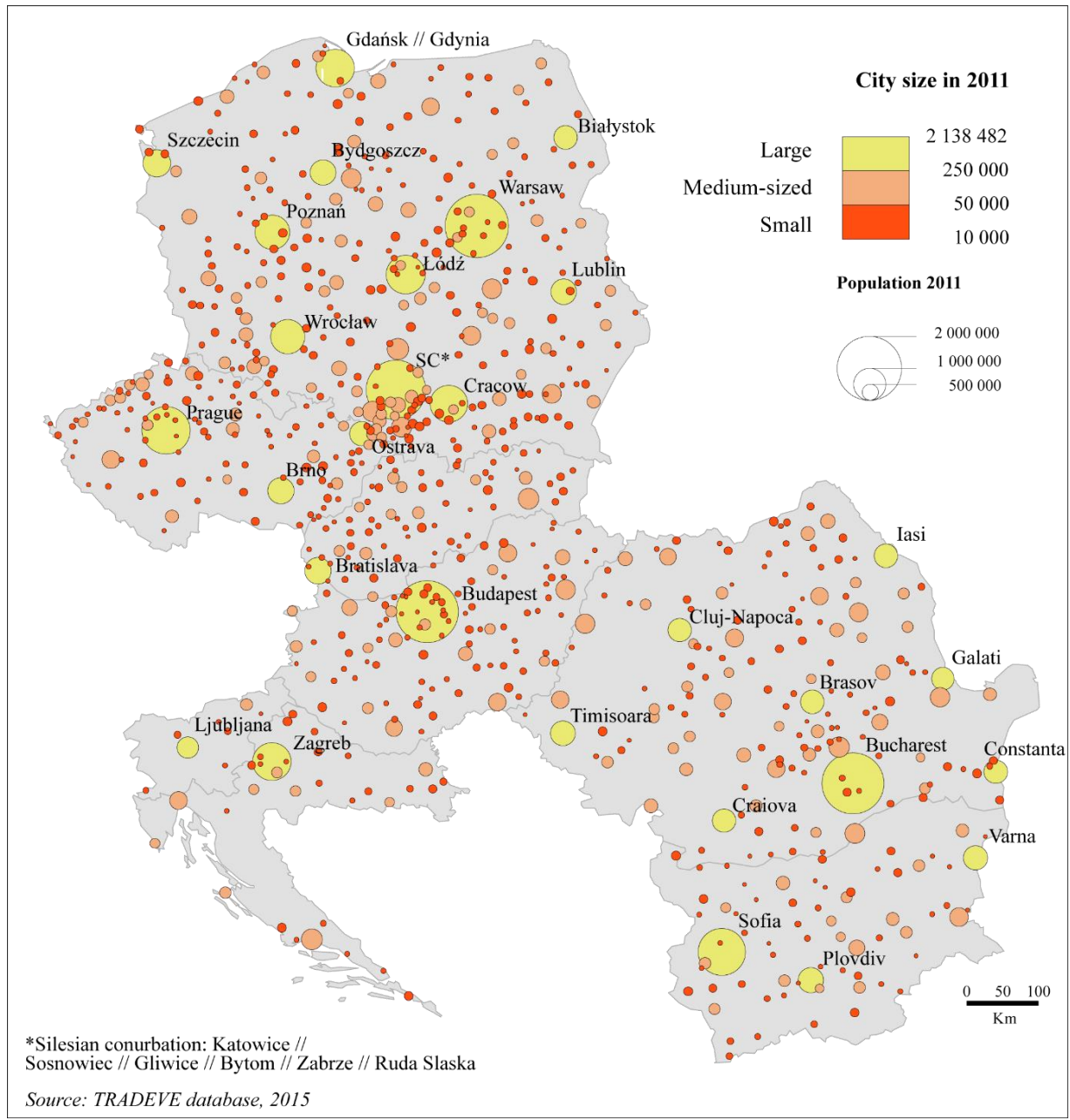

# Appendix 2. Gravity model – theoretical flows calculation[12]

$$Fij = k * \frac{MiMj}{Dij^a}$$

$F_{ij}$ is the theoretical flow between the city of origin i and the city of destination j

K is the constant indicating the overall level of mobility that allows to fix it in relation to the actual flows $k = \frac{\Sigma\ real\ flows}{\Sigma\ MiMj}$

$M_i$ et $M_j$ are the weights of the two countries, here the GDP

$D_{ij}$ is the kilometer distance between capital cities of countries

a is a constant here chosen by default and equal to 2

---

[12] The kilometre distance between cities was calculated using the proximity table tool in ArcMap. A default value of a = 2 was chosen without undertaking its adjustment.